\documentclass[journal]{IEEEtran}

\usepackage{graphicx}
\usepackage{bm}
\usepackage{enumerate}
\usepackage{float}
\usepackage{multicol}
\usepackage{color}
\usepackage{url}
\usepackage{cite}
\usepackage{amsmath}
\usepackage{amsthm}
\usepackage{amssymb}
\usepackage{array}
\usepackage{multirow}
\usepackage[table,xcdraw]{xcolor}

\usepackage{balance}

\renewenvironment{IEEEbiography}[1]
  {\IEEEbiographynophoto{#1}}
  {\endIEEEbiographynophoto}

\ifCLASSOPTIONcompsoc
  \usepackage[caption=false,font=normalsize,labelfont=sf,textfont=sf]{subfig}
\else
  \usepackage[caption=false,font=footnotesize]{subfig}
\fi

\usepackage{acronym}

\PassOptionsToPackage{hyphens}{url}\usepackage{hyperref}

\acrodef{BS}{base station}
\acrodef{BF}{beamforming}
\acrodef{RA}{receive antenna}
\acrodef{PA}{predictor antenna}
\acrodef{CDF}{cumulative distribution function}
\acrodef{DFT}{discrete Fourier transform}
\acrodef{DU}{distributed unit}
\acrodef{PDF}{probability density function}
\acrodef{CU}{centralized unit}
\acrodef{SNR}{signal-to-noise ratio}
\acrodef{CSIT}{channel state information at the transmitter}
\acrodef{HAP}{high altitude platform}
\acrodef{CSI}{channel state information}
\acrodef{bps}{bits per second}
\acrodef{VPL}{vehicular penetration loss}
\acrodef{E2E}{end-to-end}
\acrodef{MC}{mission critical}
\acrodef{MRC}{maximum ratio combining}
\acrodef{MT}{mobile terminal}
\acrodef{MRT}{maximum ratio transmission}
\acrodef{OFDM}{orthogonal frequency division multiplexing}
\acrodef{LoS}{line-of-sight}
\acrodef{NLoS}{non-line-of-sight}
\acrodef{MIMO}{multiple-input-multiple-output}
\acrodef{MRN}{moving relay node}
\acrodef{MR}{moving relay}
\acrodef{mmWave}{millimeter wave}
\acrodef{NR}{New Radio}
\acrodef{UE}{user equipment}
\acrodef{4G}{fourth generation}
\acrodef{5G}{fifth generation}
\acrodef{LTE}{Long-Term Evolution}
\acrodef{UL}{uplink}
\acrodef{DL}{downlink}
\acrodef{QoS}{quality-of-service}
\acrodef{3GPP}{3rd Generation Partnership Project}
\acrodef{TDD}{time division duplex}
\acrodef{FDD}{frequency division duplex}
\acrodef{GPS}{Global Positioning System}
\acrodef{IAB}{integrated access and backhaul}
\acrodef{SIMO}{single-input-multiple-output}
\acrodef{SISO}{single-input-single-output}
\acrodef{ZF}{zero-forcing}
\begin{document}
\captionsetup{belowskip=0pt,aboveskip=0pt}

\title{Predictor Antenna: A Technique to Boost the Performance of Moving Relays}

\author{Hao~Guo,~\IEEEmembership{Student~Member,~IEEE},
        Behrooz~Makki,~\IEEEmembership{Senior~Member,~IEEE},
        Dinh-Thuy Phan-Huy,~\IEEEmembership{Member,~IEEE},
        Erik Dahlman,
        Mohamed-Slim Alouini,~\IEEEmembership{Fellow,~IEEE},
        and Tommy~Svensson,~\IEEEmembership{Senior~Member,~IEEE}}

\maketitle

\begin{abstract}
In future wireless systems, a large number of users may access the networks via moving relays (MRs) installed on top of vehicles. One of the main challenges of MRs is rapid channel variation which may make channel estimation, and its following procedures, difficult. To address these issues,  various schemes are designed, among which predictor antenna (PA) is a candidate. The PA setup refers to a system with two (sets of) antennas on top of a vehicle, where the PA(s) positioned in front of the vehicle is(are) utilized to predict the channel state information required for data transmission to the receive antennas (RAs) aligned behind the PA. In this paper, we introduce the concept and the potentials of PA systems. Moreover, summarizing the field trials for PAs  and the 3GPP attempts on (moving) relays, we compare the performance of different PA and non-PA methods for vehicle communications in both urban and rural  areas with the PA setup backhauled through terrestrial or satellite technology, respectively. As we show, with typical parameter settings and vehicle speeds, a single-antenna PA-assisted setup can boost the backhaul throughput of MRs, compared to state-of-the-art open-loop schemes, by up to $50\%.$ 
\end{abstract}

\IEEEpeerreviewmaketitle


\section{Introduction}
Future wireless networks need to support data transmission to a large number of \acp{UE} inside the vehicles such as trains, busses, trams, and cars using high-rate applications, e.g., video streaming/sharing. One of the candidate techniques for data transmission to high-speed public transportation vehicles is \ac{MR} \cite{Yutao2013ICMmoving,Jaffry2020survey}. With an \ac{MR}, an access point installed on top of the vehicle forms its own cell inside the vehicle, and works as a relay between the  network, e.g., a terrestrial  \ac{BS}, and the in-vehicle \acp{UE}.

Compared to direct communication between the \ac{BS} and the in-vehicle \acp{UE}, the implementation of an \ac{MR} reduces the handover load/failure probability \cite{Yutao2013ICMmoving,Jaffry2020survey}, eliminates the \ac{VPL} \cite{Yutao2013ICMmoving}, and provides better propagation conditions, i.e., less path loss/shadowing with better \ac{LoS} conditions. Also, unlike typical relay nodes operating with half-duplex constraint \cite{Jaffry2020survey}, proper isolation between the indoor and outdoor antennas of an MR installed on a vehicle may give the chance for full-duplex operation, improving the spectral efficiency. With these motivations, \acp{MR} are currently of interest in, e.g., mission critical and Internet-of-vehicles, use-cases.

Indeed, the presence of \ac{CSIT} helps to provide the \acp{MR} with high data rates. However, the \ac{CSIT}-based schemes developed during 2G-4G have been  designed mostly for static, pedestrian or low speed \acp{UE}. In general, 4G systems aim to perfectly serve \acp{UE}  with speed 0-15 km/h, maintain high performance at 15-120 km/h, and assure basic services at 120-350 km/h \cite{3gpp2014requirements}. Nevertheless, various field 
trials, for example, \cite{Wu2016IEsurvey}, show considerable  performance drop of 4G systems in high speeds. This, along with other mobility issues, such as  inter-carrier interference, carrier frequency offset, and frequent handover, is partly due to the \textit{channel aging} phenomenon where with high speeds the \ac{CSIT} soon becomes inaccurate, forcing the 4G systems to fall back to \ac{CSIT}-free techniques. In such methods,   mobile \acp{UE} are provided with fairly good \ac{QoS} via diversity, i.e., by allocating, e.g., more power/spectrum resources, compared to static \acp{UE}.  This, however, may be at the cost of losing the multiplexing gain of \ac{MIMO}, which is critical for the foreseeable setups  with large antenna arrays/narrow \ac{BF}.

With this background, we need to design efficient \ac{CSIT} prediction methods for \acp{MR} for which the following alternatives have been proposed. Initially, \cite{Sternad2012WCNCWusing} showed  that  the prediction range of 0.1-0.3 carrier wavelengths in space is achievable with Kalman predictors. This is suitable to support 4G setups with low control loop delays (1-2 ms) and for pedestrians. However, it is inadequate for vehicular links at high frequencies/speeds. Indeed, one can improve the performance of Kalman prediction-based schemes at high speeds/carrier frequencies by using more frequent pilot transmissions and perform interpolation \cite{DT2015ITSMmaking}. Nevertheless, this comes with additional overhead due to the increased number of pilots, which becomes more severe in \ac{FDD} setup.  Using pre-recorded coordinate-specific \ac{CSIT} is an alternative way to increase the prediction horizon. In this case, the network requires  \acp{UE} to contribute their location information, which involves frequent data exchange and the performance may be affected by traffic variations \cite{shrika2016twc}.


\begin{figure*}
\centering
  \includegraphics[width=1.7\columnwidth]{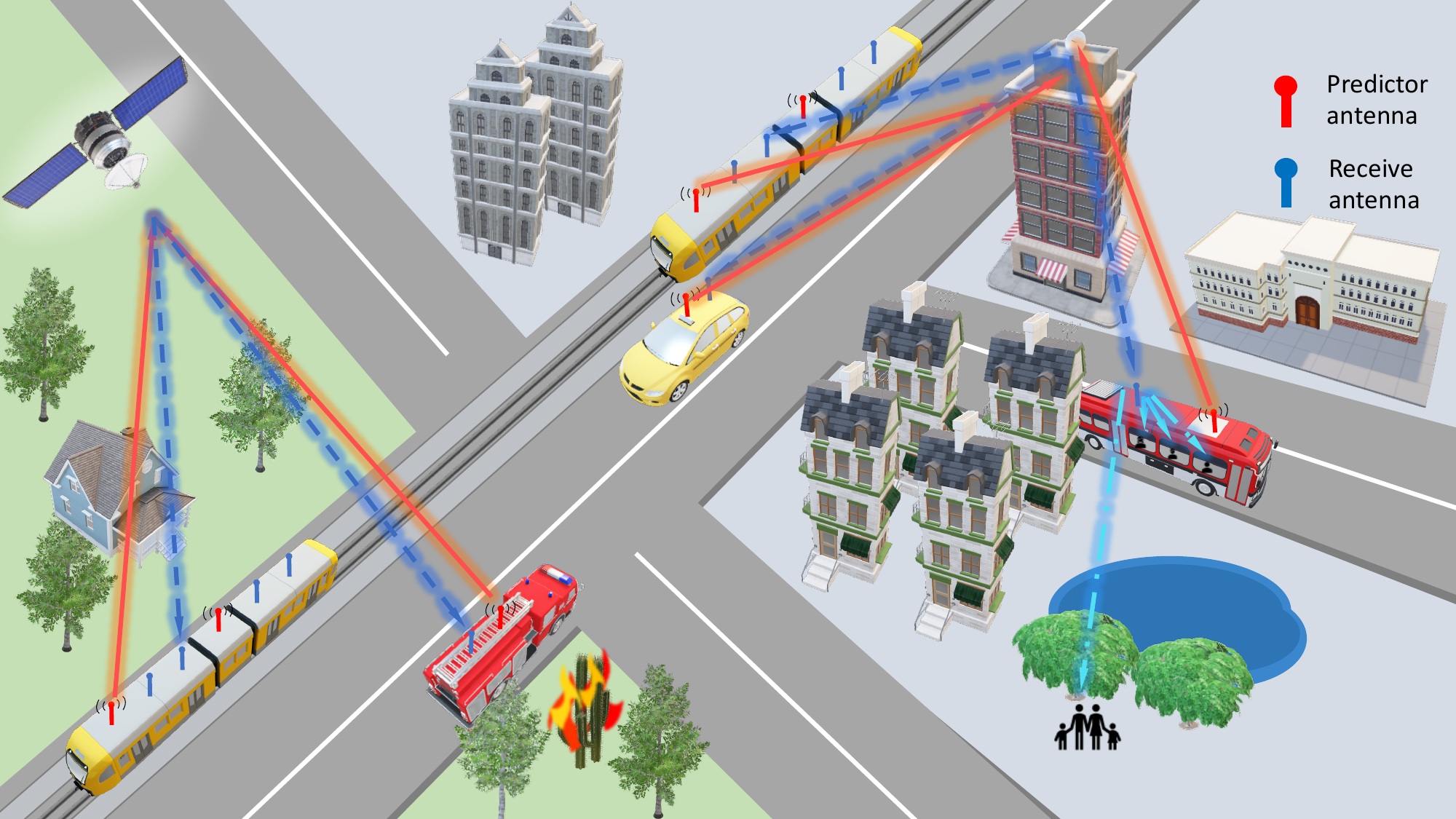}\\
\caption{PA concept in urban and rural areas with the  vehicles served by the terrestrial  \ac{BS} and satellite, respectively. }
\label{figillu}
\end{figure*}

As an alternative method to combat the channel aging phenomenon in \acp{MR}, \cite{Sternad2012WCNCWusing} proposed the \ac{PA} concept. In such a system, two groups of antennas are deployed on the roof of a vehicle, in which the \acp{PA} mounted in the front are used for predicting the channel for the \acp{RA} aligned behind the \acp{PA} once the \acp{RA} reach the same position several time slots later (see Fig. \ref{figillu}). With a TDD (T: Time) setup, at time $t$, the \ac{PA} sends the access point pilots (The \ac{PA} concept is applicable in both TDD and FDD  systems \cite{guo2020semilinear,BJ2017PIMRCpredictor}). Then, the access point, either the  \ac{BS} or the satellite/\ac{HAP}, estimates the \ac{UL} channel, uses that to obtain an estimate of  the \ac{DL} channel based on channel reciprocity, and with appropriate transmission parameter adaptation sends the data to the \ac{RA} at time $t+ L_\text{processing}$. Here, $ L_\text{processing}$ depends on the processing time at the access point, and parameter adaptation may include power, rate  and/or \ac{BF} update. In this way, as shown in Table I, compared to Kalman predictor, the presence of the \ac{PA} increases the prediction horizon considerably, where with the parameter settings of Table I and a 5 ms processing delay at the \ac{BS}, for different speeds/frequencies the \ac{PA} gives 5 times higher supported speed.  Also, prediction horizons  up to three times the  wavelengths are achievable with \ac{PA}-based system at 2.68 GHz and a velocity of 45-50 km/h, while Kalman prediction supports around 0.1-0.3 wavelengths \cite{Sternad2012WCNCWusing}.

Following \cite{Sternad2012WCNCWusing}, different testbed implementations and analytical studies were developed for \acp{PA}. Particularly, \cite{Yutao2013ICMmoving} performed system-level simulations on \ac{PA}-\ac{MR} concept and revealed that \ac{VPL} is effectively reduced. From antenna design perspective, \cite{Jamaly2014EuCAPanalysis} verified the effect of coupling compensation, and the \ac{PA} methods were practically  tested in Dresden, Germany. Other testbed-based studies include \cite{DT2015ITSMmaking} and  \cite{phan2018WSAadaptive} extending the \ac{PA} concept to \ac{MIMO} case and showing the power of channel interpolations. Also, channel measurement studies \cite{BJ2017PIMRCpredictor} verified that \ac{PA} can be used in both TDD and \ac{FDD} systems. Finally, analytical modeling of \acp{PA} were  developed in \cite{guo2020semilinear}, and various performance enhancements schemes were proposed.

In this paper, we demonstrate the potential benefits as well as the practical challenges of \acp{PA}.  We compare the performance of  PA-based schemes with different non-PA alternative methods, in terms of  \ac{E2E} throughput and prediction accuracy. The results are presented for both  urban (Section II) and rural areas (Section III) where the \ac{MR} is served by a terrestrial \ac{BS} and a satellite/\ac{HAP}, respectively. Also, we study the throughput in the presence of both  blockage and  tree foliage, and verify the effect of the \ac{BF}, estimation error and CSIT quantization  on the system performance (Section III). Finally, revisiting the field trial results on \acp{PA}, we highlight the previous and current standardization attempts on (moving) relays as well as the key points which should be addressed before the \acp{MR} and, potentially, \acp{PA} can be used in practice (Section IV). 

Our results show that the \ac{PA} concept is  an attractive candidate technique for improving the backhauling performance of \acp{MR}. Particularly, with typical parameter settings/vehicle speeds and a single PA, the implementation of the PA can increase the  backhaul throughput of the \ac{MR}, compared to open-loop systems, by $50\%$; the result that can be boosted significantly by massive \ac{MIMO}-based \ac{BF}.

\begin{table}
\centering
\caption{The maximum supported speed  in the PA- and Kalman prediction-based systems for different carrier frequencies and processing delays calculated based on \cite{Sternad2012WCNCWusing}. For the \ac{PA} setup, the antenna separation is  1.5 wavelength.}
\begin{tabular}{|c|c|c|c|c|c|}
\hline
\multicolumn{3}{|c|}{\cellcolor[HTML]{FFFFFF}Fixed delay = 5 ms}                                                                                                      & \multicolumn{3}{c|}{\cellcolor[HTML]{FFFFFF}Fixed frequency = 2.68 GHz}                                                                                          \\ \hline
                                                                            & \multicolumn{2}{c|}{\begin{tabular}[c]{@{}c@{}}Supported speed\\ (km/h)\end{tabular}} &                                                                        & \multicolumn{2}{c|}{\begin{tabular}[c]{@{}c@{}}Supported speed\\  (km/h)  \end{tabular}} \\ \cline{2-3} \cline{5-6} 
\multirow{-2}{*}{\begin{tabular}[c]{@{}c@{}}Frequency\\ (GHz)\end{tabular}} &  \cellcolor[HTML]{fc9d9a}PA               & \cellcolor[HTML]{c8c8a9}Kalman             & \multirow{-2}{*}{\begin{tabular}[c]{@{}c@{}}Delay\\ (ms)\end{tabular}} & \cellcolor[HTML]{fc9d9a}PA               & \cellcolor[HTML]{c8c8a9}Kalman             \\ \hline
1                                                                           & \cellcolor[HTML]{fc9d9a}324              & \cellcolor[HTML]{c8c8a9}65                 & 1                                                                      & \cellcolor[HTML]{fc9d9a}604              & \cellcolor[HTML]{c8c8a9}120                \\ \hline
2.68                                                                        & \cellcolor[HTML]{fc9d9a}120              & \cellcolor[HTML]{c8c8a9}24                 & 3                                                                      & \cellcolor[HTML]{fc9d9a}201              & \cellcolor[HTML]{c8c8a9}40                 \\ \hline
4                                                                           & \cellcolor[HTML]{fc9d9a}81               & \cellcolor[HTML]{c8c8a9}16                 & 5                                                                      & \cellcolor[HTML]{fc9d9a}120              & \cellcolor[HTML]{c8c8a9}24                 \\ \hline
6                                                                           & \cellcolor[HTML]{fc9d9a}54               & \cellcolor[HTML]{c8c8a9}11                 & 8                                                                      & \cellcolor[HTML]{fc9d9a}75               & \cellcolor[HTML]{c8c8a9}15                 \\ \hline
\end{tabular}
\end{table}



\section{The Potential of PA; Urab Area Study}

With vehicular speeds, the distance traveled during the typical control-loop delays of, say, $\le5$ ms, is in the order of meter or less. Hence, the moving direction can be well assumed to be almost linear. Also, as verified experimentally in \cite{Jamaly2014EuCAPanalysis}, during such a period the vehicle experiences an essentially stationary electromagnetic standing wave pattern. Then, if the \ac{RA} ends up in the same position, it will experience the same radio environment, and the \ac{CSIT} will be almost perfect (the effect of CSIT quantization and estimation error due to noisy channel measurements are studied in the following). Nevertheless, the \ac{CSIT} quality and, as a result, the overall performance may be deteriorated by spatial mismatch; if the \ac{RA} ends up in a different point from where the \ac{PA} estimated the channel, due to, e.g., the processing delay at the \ac{BS} is not the same as the time that the \ac{RA} needs to arrive to the same point as the \ac{PA}, the \ac{CSIT} will be imperfect. To address this problem,  without increasing the number of pilots, one can consider two different methods:
\begin{itemize}
\item \textbf{Adaptive-delay setup}: Knowing the vehicle speed, the transmission delay $ L_\text{processing}$ can be dynamically adapted, as a function of the antennas distance and vehicle speed, such that the \ac{RA} receives the data at the same point as the \ac{PA} sending pilots. In this case, there is potentially no spatial mismatch, at the cost of extra transmission delay. However, the delay adaptation method is applicable only for a range of vehicle speeds limited by the access point’s minimum required processing delay.  Also, cellular technologies only allow for a limited transmission time interval granularity and, in practice, there would be a residual mispointing.
\item \textbf{Nonadaptive-delay setup}: As an alternative approach, one can always consider the access point’s minimum processing delay. This method, which is more appropriate for slotted communication systems, is at the cost of possible spatial mismatch. However, we can use the spatial correlations to determine the appropriate data transmission parameters for different positions surrounding the point where the \ac{PA} was sending the pilots, and adapt the transmission parameters based on imperfect \ac{CSIT} \cite{guo2020semilinear}. 
\end{itemize}


\begin{figure}
\centering
  \includegraphics[width=1.0\columnwidth]{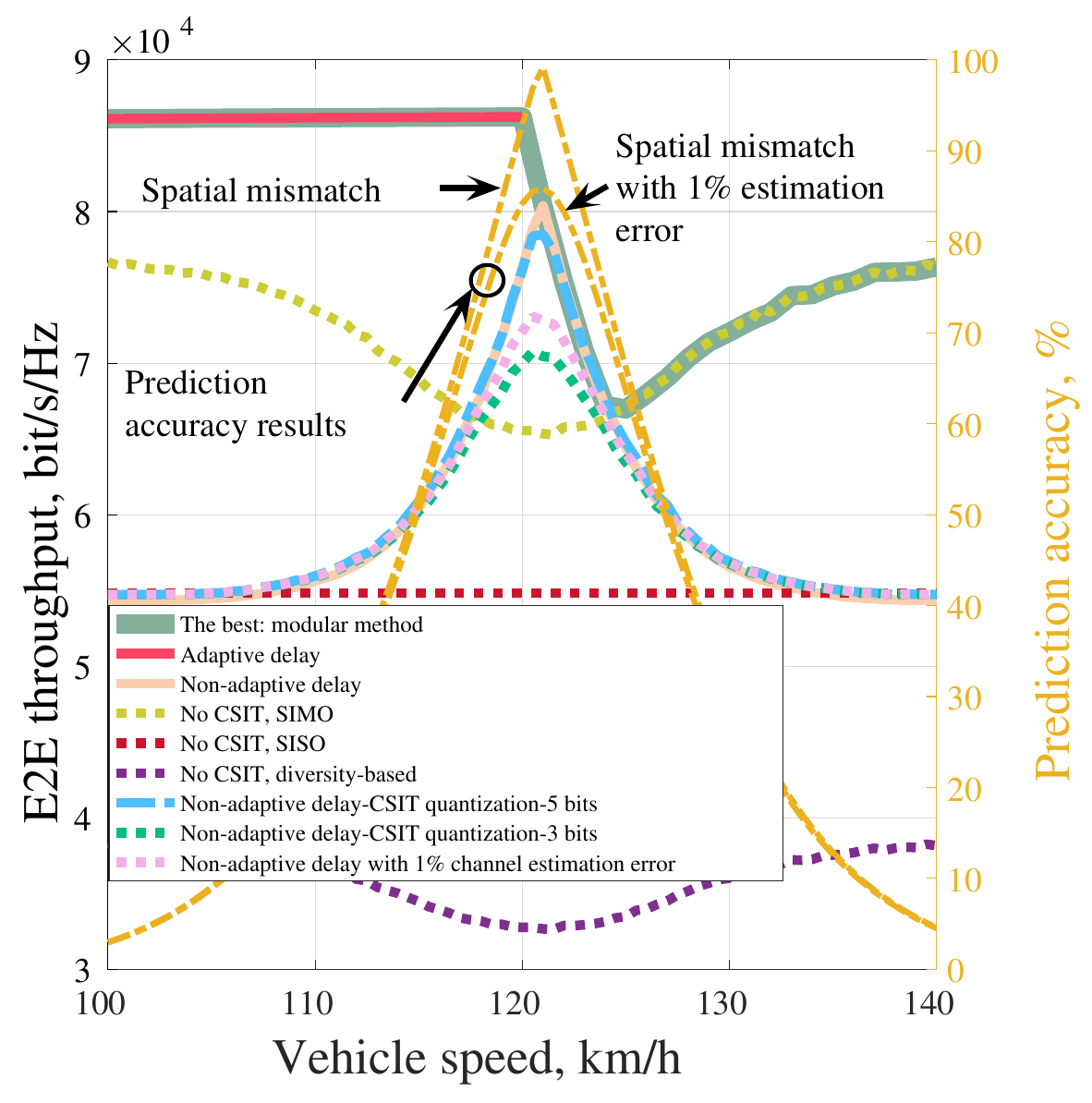}\\
\caption{\ac{E2E} throughput on the left and normalized prediction accuracy on the right for of \ac{PA} system in urban areas with spatially-correlated Rayleigh fading conditions. We set \ac{SNR} = 21 dB, codeword length = $10^4$ channel use, the minimum processing delay of the \ac{BS} =  5 ms, antenna separation =  1.5 times the wavelength, and carrier frequency = 2.68 GHz. Beside the cases with ideal channel estimation/\ac{CSIT} feedback, the  throughput for non-adaptive delay setup is given for the cases with 3, 5 CSIT quantization bit or 1\% channel estimation error due to noisy measurements. The prediction accuracy is studied for two cases: 1) effect of spatial mismatch only, and 2) effect of spatial mismatch and 1\% estimation error.}
\label{fig2}
\end{figure}


Considering an urban area with a  \ac{BS} serving the vehicle, Fig. \ref{fig2} compares the performance of the adaptive- and nonadaptive-delay \ac{PA} systems with a number of alternative schemes including:
\begin{itemize}
\item A benchmark  SISO (S: single)   scheme with single antenna on the vehicle and no \ac{CSIT} at the \ac{BS}.
\item A $1\times 2$ SIMO setup with no \ac{CSIT} at the \ac{BS}. Here, both the \ac{PA} and the \ac{RA} receive the data simultaneously from the \ac{BS}, and the receiver tries decoding by \ac{MRC} of the received signals.   
\item A diversity-based scheme where, considering a  SISO  setup, the same signal is transmitted in two spectrum resources, and an \ac{MRC}-based receiver is used for signal decoding.   
\end{itemize}
The results are presented for spatially-correlated Rayleigh-fading conditions where, using the Jakes’ correlation model and assuming uniform angular spectrum, one can model the channel around the \ac{PA} with the same procedure as in \cite[Eq. (37)-(43)]{guo2020semilinear}.

As the metric of interest, we consider  the normalized prediction accuracy and   \ac{E2E} throughput in \ac{bps}, which is defined as the average number of correctly decoded information bits per the  \ac{E2E} transmission delay. The \ac{E2E} transmission delay is given by the transmission delay plus the possible processing delay at the \ac{BS}, i.e., $L_\text{processing}$. In the adaptive-delay \ac{PA} setup the processing delay is dynamically determined based on the vehicle speed, while with a non-adaptive delay PA system $L_\text{processing}$ is set to the minimum processing delay of the \ac{BS}. In the other benchmark schemes, we have $L_\text{processing} = 0$, as the \ac{BS} is provided with no CSIT. Finally, please note that in our simulations, for each scheme, rate optimization have been utilized to maximize the \ac{E2E} throughput. In simulations, we use a 5G frame structure with 14 symbols per slot, and the total time duration is 1 ms. Here, and also in Fig. \ref{fig32}, the results are presented in bit/s/Hz, i.e., in normalized case. The results can be easily scaled depending on the considered bandwidth.

Figure \ref{fig2} demonstrates the potential of \ac{PA} system where, for a broad range of vehicle speeds, the highest \ac{E2E} throughput is achieved by the \ac{PA} method, compared to the benchmark schemes. However, there is not a single method providing the highest throughput, and a modular setup of different schemes guarantees the highest throughput for different speeds; Delay-adaptive \ac{PA} method gives the maximum throughput at low speeds, limited by the \ac{BS} minimum required processing delay. At moderate speeds, exploiting the spatial correlation and using the nonadaptive-delay \ac{PA} method leads to the highest throughput. Finally, at high speeds, where the spatial correlation between the initial position of the \ac{PA} and the final position of the \ac{RA} decreases, using both antennas for simultaneous data reception with no \ac{CSIT} at the \ac{BS} gives the maximum \ac{E2E} throughput. 

Along with spatial mismatch, different factors such as channel estimation/quantization error and mismatch of the \ac{UL} and \ac{DL} channels may affect the system performance. Figure \ref{fig2} gives the  throughput for both cases with channel estimation errors (due to noisy measurements) and quantization errors. As shown, for different parameter settings,  estimation/quantization errors affect the throughput slightly, and the PA-based scheme improves the  throughput, compared to benchmark schemes.

We concentrate on \ac{E2E} throughput, as it enables us to consider the cost of \ac{CSIT} acquisition, and compare different schemes fairly. Alternatively, as in \cite{Sternad2012WCNCWusing,DT2015ITSMmaking,Jamaly2014EuCAPanalysis,phan2018WSAadaptive,BJ2017PIMRCpredictor}, one can consider the estimation error/prediction accuracy. Figure \ref{fig2} shows the prediction accuracy of the PA setup for different speeds. As demonstrated,  the prediction accuracy results are in harmony with the \ac{E2E} throughput results, and the effect of estimation/quantization errors on the  throughput is almost negligible for a broad range of vehicle speeds.

Note that Fig. \ref{fig2} studies the system performance  for the worst-case scenario with only two antennas on the vehicle, one antenna at the \ac{BS} and no \ac{BF}. As explained in the following, with a large number of antennas the relative cost of allocating one antenna for only channel estimation decreases,  and the benefit of \ac{PA} method increases.

\subsection{MASSIVE MIMO BF IN MULTI-PATH PROPAGATION ENVIRONMENT}
\begin{figure}
\centering
  \includegraphics[width=1.0\columnwidth]{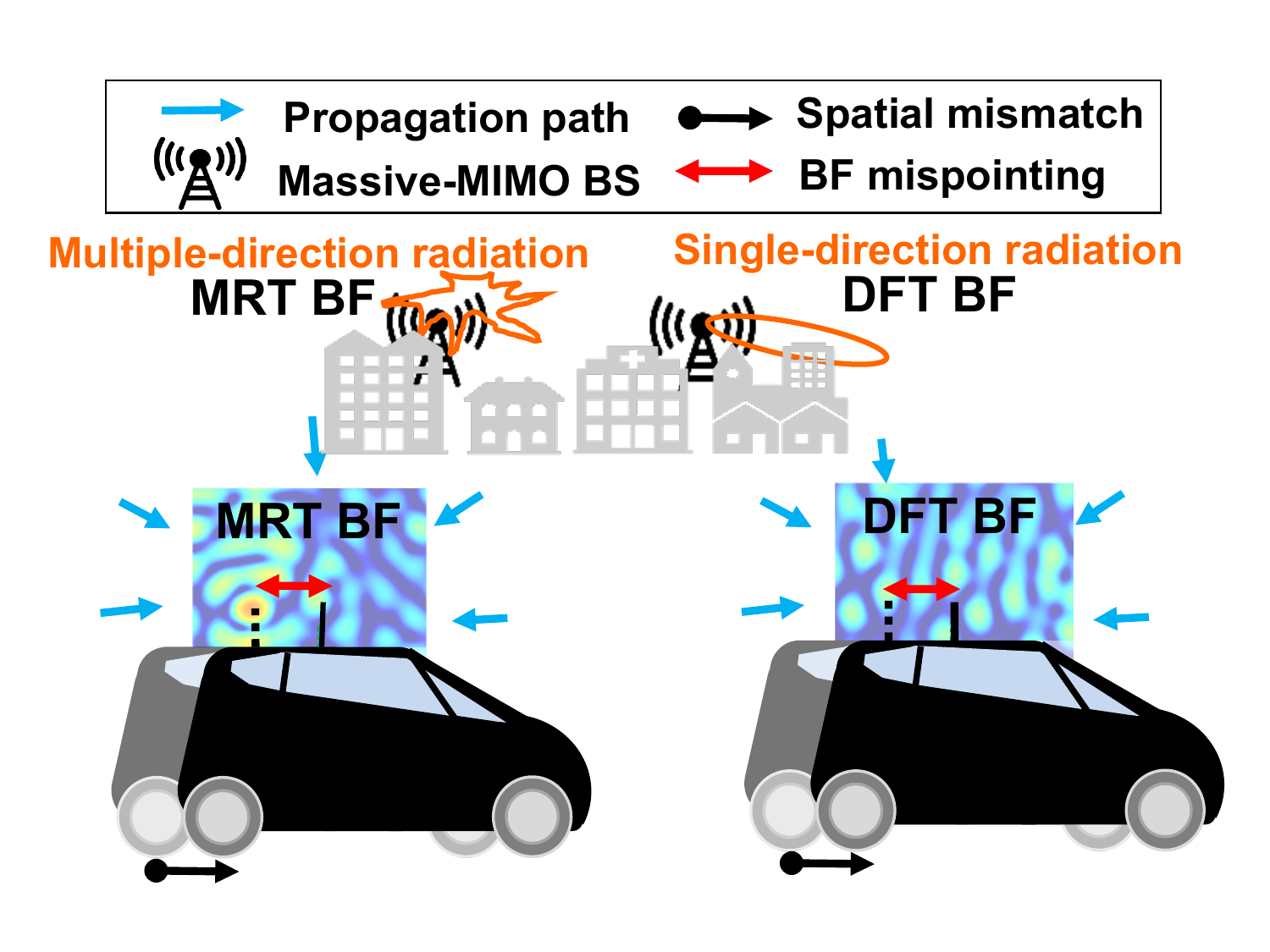}\\
\caption{Effect of spatial mismatch on the \ac{MRT}  and \ac{DFT} \ac{BF}  in an NLoS multi-path propagation environment.}
\label{figBFillu}
\end{figure}

\ac{CSIT}-based  massive \ac{MIMO} adaptive \ac{BF} schemes, e.g., \ac{MRT} and \ac{ZF}, suffer from \ac{BF} mispointing as a result of spatial mismatch, and could  benefit from the \ac{PA} \cite{DT2015ITSMmaking}. Figure  \ref{figBFillu} illustrates \ac{BF} mispointing for the \ac{MRT} and \ac{ZF} \ac{BF} in an NLoS (N: non)  multi-path propagation scenario that is likely to be encountered in an urban environment. When \ac{MRT} \ac{BF} is used in a Rayleigh two-dimensional propagation environment, the \ac{BF} pattern is close to a Bessel function, with side lobes every half wave length. Hence, even a small spatial mismatch implies a strong degradation in the received power. As illustrated in Fig. \ref{figBFillu}, \ac{DFT}-based \ac{BF}  also suffers from \ac{BF} mispointing, but in a less severe manner. Indeed, on one side, as illustrated in Fig. \ref{figBFillu}, \ac{MRT} \ac{BF} adapts to each individual path and is very sensitive to \ac{CSIT} error, whereas, \ac{DFT} \ac{BF} forms a large beam towards a single direction, and is therefore less sensitive to \ac{CSIT} error. 

\begin{figure}
\centering
  \includegraphics[width=1.0\columnwidth]{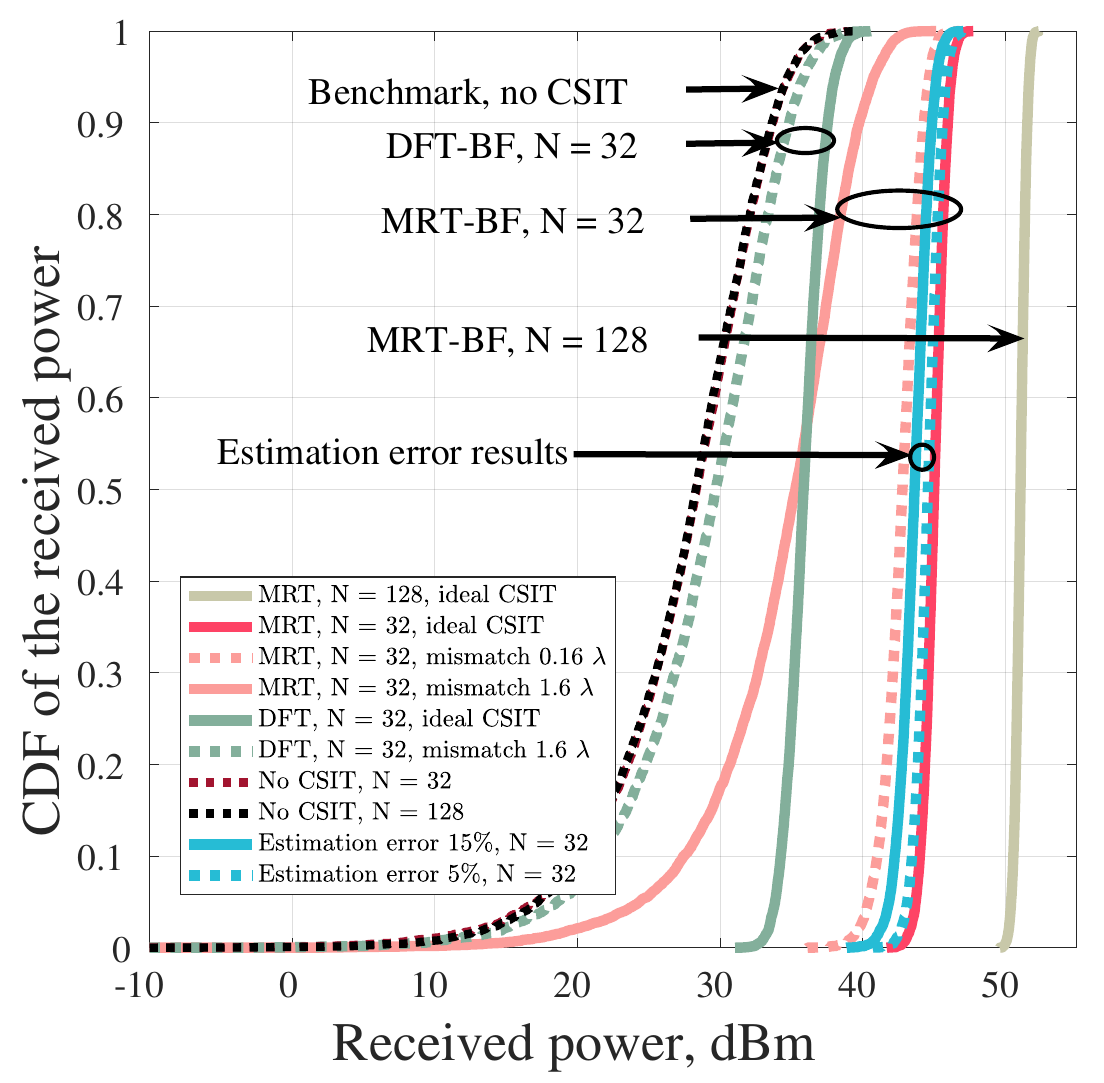}\\
\caption{Received power at the vehicle for various \ac{BF} schemes, prediction schemes (ideal prediction, without prediction, i.e., with spatial mismatch),  antenna sizes $N$, and levels of estimation error (due to noisy measurements), in a spatially-correlated Rayleigh fading environment.}
\label{figBF}
\end{figure}

Figure \ref{figBF} shows the \ac{CDF} of the received power at the vehicle side, with $N=32, 128$   antennas  at the \ac{BS},  various prediction schemes (ideal prediction and without prediction, i.e., with a spatial mismatch),  different \ac{CSIT}-based \ac{BF} schemes (\ac{MRT} and \ac{DFT}) and a scheme without \ac{CSIT} (where all antennas transmit the same signal). A typical 5G  link at 3.5 GHz carrier frequency, with  velocities of 100 km/h and 10 km/h as well as a delay of 5 ms, corresponding to a spatial mismatch of around 1.6 and 0.16 wavelength, is considered. As illustrated in Fig.  \ref{figBF}, for all schemes, the performance degradation due to spatial mismatch is in an order of magnitude.

Thanks to the \ac{PA}, \ac{BF} can be used to mitigate the spatial mispointing problem \cite{DT2015ITSMmaking}, thus improving the received power of the network, especially when the network is highly loaded with numerous moving cars or high speed trains. However, as explained, \ac{PA} alone may suffer from residual spatial mismatch when the velocity, the \ac{PA} spacing and the delay do not match. In this case, a prediction with zero residual spatial mismatch is obtained by filtering and interpolating multiple measurements that suffer from residual spatial mismatch. Recently, such schemes, with low complexity and intended for implementation and in-line running on real \ac{BS}, have been designed  \cite{BJ2017PIMRCpredictor}. Finally, experimental measurements \cite{phan2018WSAadaptive}, with a car, a 64-antenna \ac{MIMO} \ac{BS} in NLoS urban environment and  various \ac{PA} spacing values, have shown that the received power for the \ac{MRT} \ac{BF} with \ac{PA}-based prediction is close to that obtained by ideal prediction. It is also shown that both \ac{MRT} and \ac{ZF} \ac{BF}-based received powers are improved by an order of magnitude with \ac{PA}-based prediction, even when the \ac{PA} spacing, i.e., the spatial mismatch to be compensated, is as large as 3 wavelengths. Based on these studies, we can conclude that the gap between  ideal prediction and the curves with spatial mismatch, illustrated in Fig. \ref{figBF}, can be filled by \ac{PA}-based \ac{BF}. Finally, in harmony with Fig. \ref{fig2},  Fig. \ref{figBF}  shows that, with multiple antennas and \ac{BF}, the system performance  is  slightly affected by channel estimation error.

\section{Rural Area Study: PA on a Train with Many Wagons}

\begin{figure}
\centering
  \includegraphics[width=1.0\columnwidth]{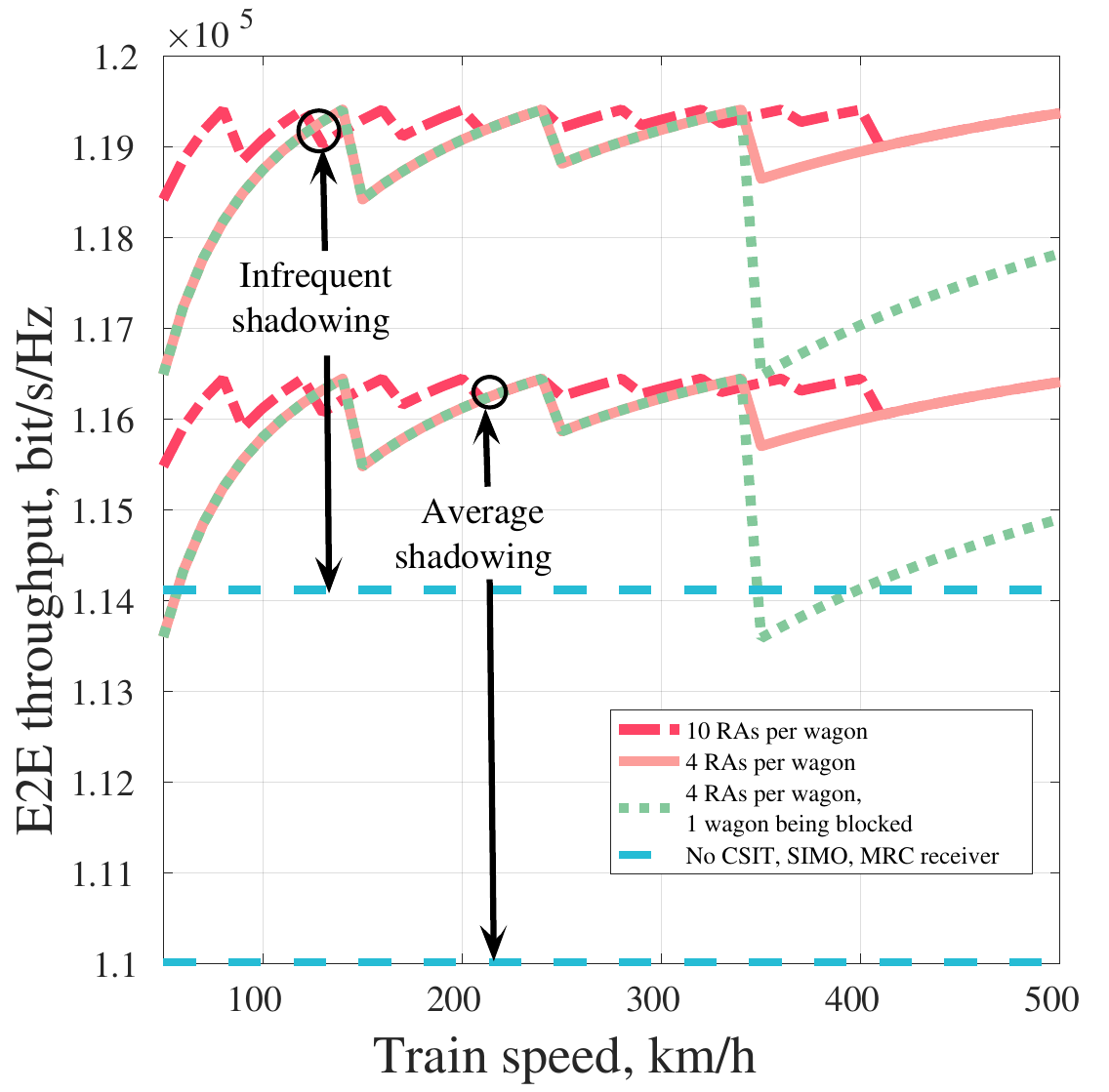}\\
\caption{\ac{E2E} throughput of the last wagon in a 10-wagon train served by the satellite,  SNR = 26 dB, codeword length = $10^4$ channel use,  minimum processing delay = 10 ms. Antenna separation between the first and the last antenna in the same wagon is 10 times the wavelength, and carrier frequency is 2.68 GHz. The distance for adjacent \ac{PA} and \ac{RA} in closest wagons is 0.2 m.}
\label{fig32}
\end{figure}


The \ac{PA} concept can be well applied in rural areas where, e.g., the high-speed train is served by a satellite in geostationary orbit or an \ac{HAP} having high geographical coverage. Particularly, the presence of multiple wagons gives the chance to implement different channel \textit{learning} and \ac{PA}-\ac{RA} pairing methods, and compensate for, e.g., blockage and channel aging. As an illustrative example, Fig. \ref{fig32} shows the \ac{E2E} throughput of the last wagon of a train with ten wagons, each equipped with a \ac{PA} and $M=4, 10$ \acp{RA}.  Here,  \textit{best-combination} scheme is used where, with an adaptive-delay method, the information of all \acp{PA} is collected, and for each wagon the \ac{PA}-\ac{RA} pair with the lowest transmission delay/highest CSIT accuracy is peaked for data transmission to the selected \ac{RA}. 

To evaluate the effect of tree foliage, the results are presented for different shadowed Rice models for land mobile satellite channels with average and infrequent light shadowings \cite[Table. III]{abdi2003new} which model the cases with moderate and low tree densities, respectively. Also, the figure verifies the robustness of the \ac{PA} system to  blockage where the  throughput is presented for the cases with one of the wagons, the ninth wagon, being fully blocked. 

The figure shows that, for a broad range of train speeds, the \ac{PA}-based scheme can boost the \ac{E2E} throughput, compared to the baseline approach with a $1\times 2$ SIMO setup using \ac{MRC} receiver, and the relative performance gain of the \ac{PA} method increases with  tree foliage. Also, the implementation of the \ac{PA}  with proper antenna pairing leads to relatively low  throughput variation at different speeds, and the throughput variation decreases with the number of \acp{RA} per wagon. Finally, although blockage leads to throughput drop in certain ranges of speed, still the \ac{PA}-based transmission is useful combating the channel aging phenomenon. Consequently, the \ac{PA} provides the in-vehicle \acp{UE} with almost constant \ac{QoS} in different environments/speeds.  Finally, while Fig. \ref{fig32} presents the simplest case with a single antenna at the satellite, with multiple antennas/wagons one can exploit the location information and dynamically adapt the \ac{BF} to reduce the effect of blockage/foliage even further.

\section{Towards Practical PAs; Standardization and Testbed Evaluations}
To validate the \ac{PA} concept, various field trials have been performed:
\begin{itemize}
    \item In 2014, we performed a field trial of the \ac{PA} system in Dresden, Germany   \cite{Jamaly2014EuCAPanalysis}. The testbed was based on installing two in-line thin $\lambda/4$ monopole antennas on the roof of a vehicle running at around 50 km/h with \ac{OFDM}, a bandwidth of 20 MHz and carrier frequency 2.68 GHz. From the field trials, the cross-correlation between the received signals of the \ac{PA} and \ac{RA} is observed to remain high ($\ge97\%$), after coupling compensation, for at least  3 times the  wavelength in both \ac{LoS} and NLoS scenarios.
    \item In 2018, our drive tests  in Stuttgart, Germany, with a massive \ac{MIMO} setup operating at 2.18 GHz showed that, at low/moderate speeds, the complex \ac{OFDM} \ac{DL} channels can be well predicted with an accuracy that enables \ac{MRT} \ac{BF} with close to ideal \ac{BF} gain  for NLoS channels \cite{phan2018WSAadaptive}.
    \item In 2018, \cite{BJ2017PIMRCpredictor} performed a testbed-based study in Dresden, Germany,  with vehicle speed of 25-50 km/h at 2.53 GHz for both \ac{LoS} and NLoS channels. The experiments validated the correlation-based analytical model, and proved that  prediction accuracy of around -13 to -7 dB is sufficient to support various transmission schemes such as precoding and spatial multiplexing.
\end{itemize}

These testbed results verify the usefulness of the \ac{PA} concept in  \acp{MR}. However, to be practically implemented, MR standardization should be first specified.

In 5G \ac{NR}, relay-based communication is mainly followed under the concept of \ac{IAB}, introduced by the \ac{3GPP} in Release 16. The goal of \ac{IAB} is to provide \ac{NR} radio-access technology not only for the access link between the \acp{UE} and the network, but also for wireless backhauling in a, possibly, multi-hop, fashion.

\ac{IAB}-like functionality based on \ac{LTE} was introduced in \ac{3GPP} Release 10 \cite{3gpp2014overview}. Also, there was a study-item on mobile relay mainly focusing on high-speed trains \cite{3gpp2014evolved}.  However, \ac{LTE}-based wireless backhaul was not extensively used, primarily due to \ac{LTE} being constrained to sub-6 GHz spectrum which is often seen as too valuable spectrum for backhauling. In contrast, \ac{NR}, in particular \ac{IAB}, can operate also in \ac{mmWave} spectrum above 10 GHz. 

In general, the \ac{IAB} architecture follows the \ac{CU}/\ac{DU} split of gNB, as introduced  in \ac{3GPP} Release 15. Here, a gNB consists of two functionally different modules:
\begin{itemize}
    \item A \ac{CU} as the unit hosting the upper layers of the gNB protocol stack.
    \item One or multiple \acp{DU} hosting  medium access control (MAC), radio link control (RLC), and physical-layer protocols.
\end{itemize}

To enable wireless backhaul, the \ac{IAB} architecture considers two different network nodes. The \ac{IAB}-donor hosts \ac{CU} and \ac{DU} functionalities, and connects to the core network via non-\ac{IAB}, e.g., fiber backhaul. The \ac{DU} of the donor may serve \acp{UE}, as a conventional gNB, but will also serve wirelessly connected \ac{IAB} nodes.

IAB nodes rely on \ac{NR} for wireless backhauling, and consist of \ac{DU} functionality that serves \acp{UE} and, possibly, child \ac{IAB} nodes in case of multi-hop \ac{IAB}. Also, an \ac{IAB} node consists of a \ac{MT} functionality which connects to the \ac{DU} of the  higher node, called the parent of the \ac{IAB} node.

In most aspects, the \ac{IAB} link, i.e., the link between MT part of an \ac{IAB} and the \ac{DU} part of its parent node operates as a conventional gNB-\ac{UE} link. As a consequence, the \ac{NR} physical, MAC, and RLC layers have limited extensions regarding the \ac{IAB}, with a major focus on the coordination of the \ac{IAB}-node MT and DUs.  For more details on \ac{IAB}, see \cite{madapatha2020integrated}. 

In principle, the MT part of an \ac{IAB} node may contain full \ac{UE} functionality, including mobility functionality. Also, it is likely to have multiple antennas at the \ac{DU} and \ac{MT} with advanced antenna/signal processing techniques. Thus, the \ac{PA} concept would be possible to implement, once the \ac{IAB}-based \ac{MR} is installed.  However, in practice the current \ac{IAB} standardization impose strong constraints on the mobility of the \ac{IAB} nodes:
\begin{itemize}
    \item Full \ac{IAB} node mobility would imply that the \ac{DU} of an \ac{IAB} node could move between different \acp{CU}, a functionality not supported by the currently standardized \ac{CU}/\ac{DU} split. 
    \item \ac{IAB}-node mobility between different parent nodes, even if they are located under the same donor, that is, the same CU,  would imply that routing tables within the \ac{IAB} nodes would have to be dynamically updated, a functionality not supported by the current \ac{IAB} specifications. 
    \item From an architecture point-of-view, nothing prevents \ac{IAB}-node mobility as long as the \ac{IAB} node remains under the same parent node. However, this would imply that the cells created by the \ac{IAB} node would not be stationary, something which would lead to many challenges in terms of cell planning and radio-resource management. 
\end{itemize}
Also, inter-node measurement, power control and interference management are challenging topics in mobile \ac{IAB}. Thus, in practice the current \ac{IAB} specifications are limited to essentially stationary \ac{IAB} nodes.  

In the early discussions on enhancements to \ac{IAB} in \ac{3GPP} Release 17, the introduction of support for mobile \ac{IAB}  was extensively discussed. However, mainly due to time limitations, it was eventually decided not to include this in the scope of the \ac{IAB} enhancements pursued in Release 17. It is not unlikely though that the introduction of mobile \ac{IAB} nodes will be further brought up again and considered for future \ac{NR} releases. In that case, we need to handle different challenges including the \ac{CSIT} accuracy given the sensitivity of the \ac{mmWave}-based narrow \ac{BF} to inaccurate \ac{CSIT}/\ac{BF} mismatch. Here, along with other methods, the \ac{PA} concept may be a useful method  potentially in combination with other alternative schemes.

\section{Conclusions}
We presented the potentials of the PA setup to improve the CSIT accuracy in high-speed vehicles. Presenting the  field trials on PA systems as well as the previous/ongoing standardization attempts on (moving) relays, we discussed the key challenges that need to be solved before the MR and, potentially, the PA setup, can be practically used. The simulation and  testbed  results show that the PA concept is a potential solution to support future adaptive antenna systems for fast-moving vehicles. However, there is still room for further theoretical and experimental research, including testbed experiments at high speeds/carrier frequencies as well as practical comparisons/combinations of various alternative methods.

\section*{Acknowledgement}
This work was supported in part by VINNOVA (Swedish Government Agency for Innovation Systems) within the VINN Excellence Center ChaseOn. Thank Yigeng Zhang from University of Huston for the help on illustrations.

\bibliographystyle{IEEEtran}

\bibliography{main.bib}


\begin{IEEEbiography}{Hao Guo} [S'17] (hao.guo@chalmers.se)  is currently pursuing his PhD degree with Department of Electrical Engineering, Chalmers, Sweden.
\end{IEEEbiography}

\begin{IEEEbiography}{Behrooz Makki} [M'19, SM'19]  works as Senior Researcher in Ericsson Research, Sweden. 

\end{IEEEbiography}

\begin{IEEEbiography}{Dinh-Thuy Phan-Huy} currently works as Project Manager and Senior R\&D Engineer on wireless communications at Orange, France. 
 
\end{IEEEbiography}

\begin{IEEEbiography}{Erik Dahlman} is currently Senior Expert in Radio Access Technologies within Ericsson Research. 
\end{IEEEbiography}

\begin{IEEEbiography}{Mohamed-Slim Alouini} 
[S'94, M'98, SM'03, F'09]  is a Professor of Electrical Engineering in King Abdullah University of Science and Technology, Thuwal, Saudi Arabia.
\end{IEEEbiography}

\begin{IEEEbiography}{Tommy Svensson} [S’98, M’03, SM’10]  is a Professor in Communication Systems at Chalmers University of Technology, Sweden. 
\end{IEEEbiography}
\end{document}